\documentstyle[11pt]{article}
\newcommand{\quattrova}{($\phi^{\scriptscriptstyle a},
c^{\scriptscriptstyle a},
\lambda_{\scriptscriptstyle  a},{\bar c}_{\scriptscriptstyle a}$)~}

\newcommand{\be}{\begin{equation}}
\newcommand{\ee}{\end{equation}}
\newcommand{\bea}{\begin{eqnarray}}
\newcommand{\eea}{\end{eqnarray}}

\def \HT{{\widetilde{\mathcal H}}}

\def \KT{{\widetilde{\mathcal K}}}
\def \DT{{\widetilde{\mathcal D}}}

\def \HCT{{\widehat{\widetilde{\mathcal H}}}}
\def \KCT{{\widehat{\widetilde{\mathcal K}}}}
\def \DCT{{\widehat{\widetilde{\mathcal D}}}}

\def \QH{{Q_{\scriptscriptstyle H}}}
\def \QK{{Q_{\scriptscriptstyle K}}}
\def \QD{{Q_{\scriptscriptstyle D}}}

\def \QBH{{\overline{Q}}_{\scriptscriptstyle H}}
\def \QBK{{\overline{Q}}_{\scriptscriptstyle K}}
\def \QBD{{\overline{Q}}_{\scriptscriptstyle D}}
\def \Qb{ Q_{\scriptscriptstyle BRS}}
\def \QBb{{\overline {Q}}_{\scriptscriptstyle BRS}}

\def \QCH{{\widehat{Q}_{\scriptscriptstyle H}}}

\def \QCBH{{\widehat{\overline{Q}}}_{\scriptscriptstyle H}}

\def \QCb{{\widehat Q}_{\scriptscriptstyle BRS}}
\def \QCBb{{\widehat{\overline {Q}}}_{\scriptscriptstyle BRS}}

\def \NH{ N_{\scriptscriptstyle H}}

\def \HS{H_{\scriptscriptstyle SUSY}}

\newcommand{\scite}{~\cite}

\begin{document}

\baselineskip =15.5pt
\pagestyle{plain}
\setcounter{page}{1}

\vfil

\begin{center}
{\huge A New Supersymmetric Extension of Conformal Mechanics}
\end{center}

\vspace{1cm}

\begin{center}
{\large E.Deotto, G.Furlan, E.Gozzi}\\
\vspace {1mm}
Dipartimento di Fisica Teorica, Universit\`a di Trieste, \\
Strada Costiera 11, P.O.Box 586, Trieste, Italy \\ and INFN, Sezione 
di Trieste.\\
\vspace {1mm}
\vspace{3mm}
\end{center}

\vspace{1cm}

\begin{abstract}

\noindent
In this paper a new supersymmetric extension of conformal mechanics is put
forward. The beauty of this extension is that all variables have a clear
geometrical meaning and the super-Hamiltonian  turns out to be the
Lie-derivative of the Hamiltonian flow of standard conformal mechanics. 
In this paper we also provide a supersymmetric extension of the other conformal
generators of the theory and find  their ``square-roots". 
The whole superalgebra of these charges is then analyzed in details. 
We conclude the paper  by showing that, using superfields,  
a constraint can be built which provides the exact solution 
of the  system. 

\end{abstract}
\vspace{1cm}

%%%%%%%%%%%%%%%%%%%%%%%%%%%%%%%%%%%%%%%%%%%%%%%%%%%%%%%
%%%%%%%%%%%%%%%%%%%%%%%%%%%%%
\section{Introduction}
In 1976 a conformally-invariant quantum mechanical model
was proposed and solved in ref. [1]. New interest in the model
has been recently generated by the discovery \cite{KAL} 
that the dynamics of a particle near the horizon 
of an extreme Reissner-Nordstr\o m black-hole is governed in its radial motion
by the Lagrangian of ref. \cite{DFF}. A supersymmetric
extension of conformal mechanics was proposed later by two 
independent  groups\scite{FUB} and also  the supersymmetric version
seems to hold some interest for black-hole physics.

In this paper we shall put forward a new supersymmetric extension of conformal
mechanics. It is based on a path-integral approach to classical mechanics
developed in ref.\scite{ENNIO}.  The difference between our extension and the
one of ref.\scite{FUB}, which was tailored on the supersymmetric quantum
mechanics of Witten\scite{WITTE}, is that the authors of ref.\scite{FUB} took 
the original conformal Hamiltonian and added a Grassmannian part in order to 
make the whole Hamiltonian supersymmetric. Our procedure and extension is different and more 
geometrical 
as will be explained later on in the paper. 

The paper is organized in the following manner: In section {\bf 2} we give a 
very brief outline of conformal mechanics\scite{DFF} and of the
supersymmetric extension present in the literature\scite{FUB}; 
in section {\bf 3} we put forward our 
supersymmetric extension and explain its geometrical
structure. In the same section  we build a whole set of charges connected 
with our extension and study their algebra in detail. 
In section {\bf 4} we show that, differently
from the superconformal algebra of\scite{FUB} where the even part had
a spinorial representation on the odd part, ours is a non-simple
superalgebra whose even part has a reducible and  integer representation 
on the odd part. In section {\bf 5} we give a superspace version for our 
model and, like for the old conformal mechanics, we  provide an exact solution of 
the model. This is given by solving a  constraint 
in superfield space. Details of the calculations
can be found in a longer version of this paper\scite{PINO}.
%%%%%%%%%%%%%%%%%%%%%%%%%%%%%%%%%%%%%%%%%%%%%%%%%%%%%%%
%%%%%%%%%%%%%%%%%%%%%%
\section{The Old Conformal and Supersymmetric Extended Mechanics.}

The Lagrangian for conformal mechanics proposed in\scite{DFF}  is

\be
\label{eq:uno}
L=\frac{1}{2}\left[{\dot q}^{2}-{g\over q^{2}}\right].
\ee

\noindent This Lagrangian is invariant under the
following transformations:

\be
\label{eq:due}
t^{\prime}={{\alpha t+\beta}\over {\gamma t + \delta}};~~~
q^{\prime}(t^{\prime})= {q(t)\over (\gamma t +\delta)};~~
\mbox{with}~~\alpha\delta-\beta\gamma=1,
\ee

\noindent which are nothing else than the conformal transformations in 0+1 dimensions.
They are made of the combinations of the following three transformations:
\bea
\label{eq:tre}
t^{\prime} & = &\alpha^{2}t~~~~~~~\mbox{\it dilations},\\
\label{eq:quattro}
t^{\prime} & = & t+\beta ~~~~~\mbox{\it time-translations},\\
\label{eq:cinque}
t^{\prime} & = & {t\over {\gamma t+1}}~~~ \mbox{\it special-conformal~transformations.}
\eea

\noindent The associated Noether charges \scite{DFF} are:

\be
\label{eq:sei}
H=\frac{1}{2}\left(p^{2}+{g\over q^{2}}\right);
~~D=tH-{1\over 4}(qp+pq);~~
K=t^{2}H-\frac{1}{2}t(qp+pq)+\frac{1}{2}q^{2}.
\ee

\noindent The quantum algebra of the three Noether
charges above is:

\be
\label{eq:sette}
[H,D] = iH;~~
[K,D]=-iK;~~
[H,K]=2iD.
\ee

\noindent  The $H,D,K$
above are explicitly dependent on $t$, but they are conserved, i.e.: 

\be
\label{eq:otto}
{\partial D\over \partial t}\neq 0;~~{\partial K\over\partial t}\neq 0;~~
\frac{dD}{dt}=\frac{dK}{dt}=0.
\ee

\noindent As the $H,D,K$ are conserved, their expressions at $t=0$ which are
\be
\label{eq:nove}
H_{0}=\frac{1}{2}\left[p^{2}+{g\over q^{2}}\right];~~
D_{0}= -{1\over 4}\left[qp+pq\right];~~
K_{0}=\frac{1}{2}q^{2}
\ee

\noindent satisfy the same algebra as those at time $t$.

The supersymmetric extension of this model,
proposed in ref.\scite{FUB}, has the following Hamiltonian:

\be
\label{eq:dieci}
\HS = \frac{1}{2}\left(p^{2}+{g\over q^{2}}+{\sqrt {g}\over
q^{2}}[\psi^{\dag},\psi]_{\scriptscriptstyle -}\right),
\ee
\noindent
where $\psi,\psi^{\dag}$ are Grassmannian variables whose anticommutator
is $[\psi,\psi^{\dag}]_{\scriptscriptstyle +}=1$. In $\HS$
there is a first bosonic piece which is the conformal Hamiltonian of 
eq.(\ref{eq:uno}), plus a Grassmannian part. Note that the equations
of motion for ``$q$" have an extra piece with respect to the equations
of motion of the old conformal mechanics\scite{DFF}.

$\HS$ can be written as a particular form of
supersymmetric quantum mechanics\scite{WITTE}:

\be
\label{eq:undici}
\HS=\frac{1}{2}\left[Q,Q^{\dag}\right]_{\scriptscriptstyle +}=
\frac{1}{2}\left(p^{2}+\left({dW\over dq}\right)^{2}-[\psi^{\dag},\psi
]_{\scriptscriptstyle -}{d^{2}W\over dq^{2}}\right)
\ee

\noindent where the  supersymmetry charges are given by:

\be
\label{eq:dodici}
Q=\psi^{\dag}\left(-ip+{dW\over dq}\right);~~
Q^{\dag}=\psi\left(ip+{dW\over dq}\right);
\ee

\noindent and where  $W$ is  the superpotential which, in this case of conformal mechanics,
turns out to be:
\be
\label{eq:tredici}
W(q)=\sqrt{g}\log q.
\ee

\noindent If we perform a supersymmetric
transformation combined with a conformal one generated by the $(H,K,D)$,
we get what is called a {\it superconformal} transformation. 
In order to understand  this better let us list the following eight operators:
\bea
\label{eq:quattordici}
&&H=\displaystyle\frac{1}{2}\Bigl[p^{2}+{{g+2\sqrt{g}B}\over q^{2}}\Bigr];~~~~~~
D = \displaystyle -{[q,p]_{\scriptscriptstyle +}\over 4};~~~~~~
K = \displaystyle{q^{2}\over 2};\\
\label{eq:quindici}
&&B=\displaystyle{[\psi^{\dag},\psi]_{\scriptscriptstyle -}\over 2};~~~~~
Q = \displaystyle \psi^{\dag}\Bigl[-ip+{\sqrt{g}\over q}\Bigr];~~~~~
Q^{\dag} =\displaystyle \psi\Bigl[ip+{\sqrt{g}\over q}\Bigr];\\
\label{eq:sedici}
&&S=\displaystyle \psi^{\dag}q;~~~~~~~~~~~~~~~~~
S^{\dag}=\displaystyle \psi q.
\eea

The algebra of these operators is closed and given in the table below:

\begin{center}
{\bf TABLE 1}
\end{center}
\[
\begin{array}{|lll|}
\hline & & \\

[H,D]=iH; \hspace{3cm}&[K,D]=-iK; \hspace{3cm} &[H,K]=2iD; \\

[Q,H]=0; &[Q^{\dag},H]=0; &[Q,D]={i\over 2}Q; \\

[Q^{\dag},K]=S^{\dag}; &[Q,K]=-S; &[Q^{\dag},D]={i\over 2}Q^{\dag}; \\

[S,K]=0; &[S^{\dag},K]=0; &[S,D]=-{i\over 2}S; \\

[S^{\dag},D]=-{i\over 2}S^{\dag}; &[S,H]=-Q; &[S^{\dag},H]=Q^{\dag}; \\

[Q,Q^{\dag}]=2H; &[S,S^{\dag}]=2K; & \\

[B,S]=S; &[B,S^{\dag}]=-S^{\dag}; & \\
 
[Q,S^{\dag}]=\sqrt{g}-B+2iD; &[B,Q]=Q; &[B,Q^{\dag}]=-Q^{\dag}. \\

& & \\
\hline
\end{array} 
\]

\noindent 
The square-brackets $[(.),(.)]$ in the algebra above 
are  {\it graded}-commutators.
We notice that the commutators of the supersymmetry 
generators $ (Q,Q^{\dag}) $ with the three conformal generators 
$ (H,K,D) $ generate
a new operator which is $ S $. Including this new one we generate an algebra
which is closed  provided that we introduce the operator $B$ of eq.(15) 
which is the last operator we need. 

%%%%%%%%%%%%%%%%%%%%%%%%%%%%%%%%%%%%%%%%%%%%%%%%%%%%%%%
%%%%%%%%%%%%%%%%%%%%%%%%%%%%%
\section{A New Supersymmetric Extension.}

In this section we are going to present a new supersymmetric extension
of conformal mechanics. This extension is based on a path-integral
approach to classical mechanics (CM) developed in ref.\scite{ENNIO}.
Let us start with a system living on a $2n$-dimensional phase space ${\cal M}$ 
---whose coordinates we indicate as $\phi^{a}$ with $a=1,\ldots,2n$, i.e.: $\phi^{a}=(q^1,\ldots,
q^n;p^1,\ldots,p^n)$---and having an Hamiltonian~ $H(\phi)$. 
The equations of motion are then:
~${\dot\phi }^{a}=\omega^{ab}{\partial H\over\partial\phi ^{b}}$ where 
$\omega^{ab}$ is the usual symplectic matrix.
\noindent By {\it path integral for CM} we mean a functional integral that forces all paths
in ${\cal M}$ to sit on the classical ones. The {\it classical} analog of the quantum generating 
functional is then:
\[
Z_{\scriptscriptstyle CM}[J]=N\int{\cal D}\phi~\tilde{\delta}[\phi
(t)-\phi _{cl}(t)]\exp\left[\int J\phi~dt\right],
\]
\noindent where $\phi$ are the $\phi^{a}\in{\cal M}$, $\phi_{cl}$ are the
solutions of the equations of motion.
$J$ is an external current and $\widetilde{\delta}[.]$ is a functional
Dirac delta which forces
every path $\phi(t)$ to sit on a classical ones $\phi_{cl}(t)$. 
Let us first
rewrite the functional Dirac delta in the $Z_{\scriptscriptstyle CM}$ above as:
\[
{\tilde\delta}[\phi -\phi _{cl}]={\tilde\delta}[{\dot\phi
^{a}-\omega^{ab}
\partial_{b}H]~det [\delta^{a}_{b}\partial_{t}-\omega^{ac}\partial_{c}\partial
_{b}H}].
\]
The next step is to  write the $\tilde{\delta}[.]$
as a Fourier transform over some new variables $\lambda_{a}$
and to exponentiate the determinant via
Grassmannian variables $\bar{c}_a, c^{a}$. The final result is
\[
Z_{\scriptscriptstyle CM}[0]=\int{\cal D}\phi ^{a}{\cal D}\lambda_{a}{\cal
D}c^{a}{\cal D}
{\bar c}_{a}\exp\left[i\int dt{\widetilde{\cal L}}\right]~~
\mbox{with } ~~{\widetilde{\cal L}}=\lambda_{a}[{\dot\phi }^{a}-\omega^{ab}\partial_{b}H]+
i{\bar c}_{a}[\delta^{a}_{b}\partial_{t}-\omega^{ac}\partial_{c}\partial_{b}H]
c^{b}.
\]
One can derive the equations of motion from this Lagrangian which, 
for $\phi^{a}$,
are the standard Newton equations while for $c^{a}$ are the equations of
the first variations. This last thing allow us to identify the $c^{a}$
with the basis of the forms $d\phi^{a}$\scite{MARS}.

The equations of motions can be derived\scite{ENNIO} also from the Hamiltonian
associated to the Lagrangian above which is
\begin{equation}
\label{eq:diciassette}
\widetilde{\cal
H}=\lambda_a\omega^{ab}\partial_bH+i\bar{c}_a\omega^{ac}(\partial_c\partial_bH)c^b.
\end{equation} 

\noindent As the $Z_{\scriptscriptstyle CM}$ is a path integral we can also define 
the concept of {\it commutator}
as Feynman did in the quantum case. The result\scite{ENNIO} is:
$\langle[\phi ^{a},\lambda_{b}]\rangle=i\delta^{a}_{b};~~~~\langle[{\bar
c}_{b},
c^{a}]\rangle=\delta^{a}_{b}$. 

Using this operatorial formulation and the fact that the $c^{a}$ can be
identified with forms, it is easy to prove that $\HT$ is nothing else than
the Lie-derivative of the Hamiltonian flow\scite{ENNIO} generated by
$H$.
The reader may remember that the concept
of Lie-derivative was mentioned also in the second of refs.\scite{WITTE}.
There anyhow the connection between Lie-derivative and Hamiltonian
was not as direct as here. Moreover the Lie-derivative was not linked
to the flow generated by the conformal potential (1) but with the flow
generated by  the superpotential (13).

The Hamiltonian $\HT$ has various {\it universal} symmetries\scite{ENNIO}
all of which have been studied geometrically\scite{ENNIO}\scite{MARMAU}. 
The associated charges are:
\bea
\label{eq:diciotto}
&&Q_{\scriptscriptstyle BRS}=\displaystyle i c^{a}\lambda_{a};~~~~~
{\overline Q}_{\scriptscriptstyle BRS}=\displaystyle i {\bar c}_{a}\omega^{ab}
\lambda_{b};\\
&&Q_{g}=\displaystyle c^{a}{\bar c}_{a};~~~~
C =\displaystyle {\omega_{ab}c^{a}c^{b}\over 2};~~~~
{\overline C}=\displaystyle{\omega^{ab}{\bar c}_{a}{\bar c}_{b}\over 2}\nonumber \\
&&\NH=\displaystyle c^{a}\partial_a H;~~~~~
{\overline \NH}=\displaystyle {\bar c}_{a}\omega^{ab}\partial_{b}H.\nonumber
\eea

\noindent Using the correspondence between Grassmannian variables and forms,
the $\Qb$ turns out to be nothing else\scite{ENNIO} than the exterior 
derivative on phase-space.\scite{MARS}. The $Q_{g}$, or ghost charge, is the 
form-number which  is always conserved by the Lie-derivative. Similar geometrical meanings
can be found\scite{ENNIO} for the other charges that are listed above.
Of course linear combinations of them are also conserved and there are two
combinations which deserve our attention. They are the following charges:

\be
\label{eq:diciannove}
\QH\equiv\Qb-\beta\NH;~~~~~~~~~~\QBH\equiv\QBb+\beta{\overline\NH};
\ee

\noindent (where $\beta$ is an arbitrary dimensionful parameter) 
which are true supersymmetry charges because, besides commuting
with $\HT$, they give:
$[\QH,\QBH]=2i\beta\HT$.~~{\it This proves that our }$\HT$ {\it is supersymmetric}. 
To be precise it is an $N=2$
supersymmetry. One realizes immediately that $H$ acts
as a sort of superpotential for the supersymmetric Hamiltonian $\HT$. 
All this basically means that  we can obtain a  supersymmetric Hamiltonian $\HT$~out of any
system with Hamiltonian $H$~and, moreover , our $\HT$ has a nice
geometrical meaning being the Lie-derivative of the Hamiltonian
flow generated by $H$. 

We will now build the $\HT$ of the conformal invariant system given
by the Hamiltonian of eq.(\ref{eq:sei}), that means we insert
the $H$ of eq.(\ref{eq:sei}) into the $\HT$ of eq.(\ref{eq:diciassette}).
The result is:
\be
\label{eq:venti}
\HT=\lambda_q p+\lambda_p\frac{g}{q^3}
+i\bar{c}_qc^p-3i\bar{c}_pc^q\frac{g}{q^4},
\ee 
\noindent where the indices $(.)^{q}$ and $(.)^{p}$ on the variables $(\lambda, c, {\bar c})$ replace
the indices $(.)^{a}$ which appeared in the general formalism because here we have only one degree of
freedom. The two supersymmetric 
charges of eq.(\ref{eq:diciannove}) are in this case
\be
\label{eq:ventuno}
\QH=\Qb+\beta\left(\frac{g}{q^3}c^q-pc^p\right);~~~~
\QBH=\QBb+\beta\left(\frac{g}{q^3}\bar{c}_p+p\bar{c}_q\right). 
\ee

\noindent It was one of the central points of the original paper\scite{DFF} on conformal
mechanics that the Hamiltonians of the system could be, beside $H_{0}$
of eq.(\ref{eq:nove}), also $D_{0}$ or $K_{0}$ 
or any linear combination of them. In the same manner as we built the
Lie-derivative~$\HT$~associated to $H_{0}$, we can also
build the Lie-derivatives associated to the flow generated by $D_{0}$
and $K_{0}$. We just have to insert $D_{0}$ or $K_{0}$ in place of $H$
as superpotential in the $\HT$ of eq.(\ref{eq:diciassette}). Calling the
associated Lie-derivatives as $\DT_0$ and 
$\KT_0$, what we get is:

\be
\label{eq:ventidue}
\DT_0=\frac{1}{2}[\lambda_pp-\lambda_qq+i(\bar{c}_pc^p-\bar{c}_qc^q)];~~ 
\KT_0=-\lambda_pq-i\bar{c}_pc^q. 
\ee

\noindent Both $\DT_0$ and $\KT_0$ are ``supersymmetric" in the sense that 
there are the ``square" of some charges:
$[\QD,\QBD]=4i\gamma\DT_0;~~
[\QK,\QBK]=2i\alpha\KT_0$, 
given by
\bea
\label{eq:ventitre}
&&\QD=\Qb+\gamma(qc^p+pc^q);~~~~~
\QBD=\QBb+\gamma(p\bar{c}_p-q\bar{c}_q);\\
\label{eq:ventiquattro}
&&\QK=\Qb-\alpha qc^q;~~~~~~~~~
\QBK=\QBb-\alpha q\bar{c}_p.
\eea
\noindent The $\alpha$ and $\gamma$ are arbitrary constant variables like $\beta$ was.
Let us list all the operators we have found so far:
\begin{center}
{\bf TABLE 2}
\end{center}

\[
\begin{array}{|lcl|}
\hline & & \\
\HT=\displaystyle\lambda_q p+\lambda_p\frac{g}{q^3}+i\bar{c}_q
c^p-3i\bar{c}_pc^q\frac{g}{q^4}; & &H=\displaystyle\frac{1}{2}\left(p^2+\frac{g}{q^2}\right);\\
\KT_0=\displaystyle-\lambda_pq-i\bar{c}_pc^q; && K_0=\frac{1}{2}q^2;\\
\DT_0=\displaystyle\frac{1}{2}[\lambda_pp-\lambda_qq+i(\bar{c}_pc^p-\bar{c}_qc^q)];&&D_0=\displaystyle
-
\frac{1}{2}qp;\\
\Qb= \displaystyle i(\lambda_qc^q+\lambda_pc^p);&&\QBb= \displaystyle 
i(\lambda_p\bar{c}_q-\lambda_q\bar{c}_p);\\
\QH=\displaystyle\Qb+\beta\left(\frac{g}{q^3}c^q-pc^p\right);
&&\QBH=\displaystyle\QBb+\beta\left(\frac{g}{q^3}\bar{c}_p+p\bar{c}_q\right); \\
\QK=\displaystyle\Qb-\alpha qc^q; &&\QBK=\displaystyle\QBb-\alpha q\bar{c}_p;\\
\QD=\displaystyle\Qb+\gamma(qc^p+pc^q);&&\QBD=\displaystyle\QBb+\gamma(p\bar{c}_p-q\bar{c}_q).\\
&&\\
\hline
\end{array}
\]
We find that they are the minimum number in order to make a closed algebra
which is  written in the TABLE  below:
\begin{center}
{\bf TABLE 3}
\end{center}
\[
\hspace{-0.5cm}
\begin{array}{|lll|}

\hline 

& & \\

[\HT,\DT_0]=i\HT; \hspace{3cm}&[\KT,\DT_0]=-i\KT_0; \hspace{3cm}  &[\HT,\KT_0]=2i\DT_0; \\

[\QH,\HT]=0; &[\QBH,\HT]=0; &[\QH,\QBH]=2i\beta\HT; \\

[\QH,\DT_0]=i(\QH-\Qb); &[\QBH,\DT_0]=i(\QBH-\QBb); & \\

[\QH,\KT_0]=i\beta\gamma^{-1}(\QD-\Qb); &[\QBH,\KT_0]=i\beta\gamma^{-1}(\QBD-\QBb); & \\

[\Qb,\HT]=[\QBb,\HT]=0; & [\Qb,\KT]=[\QBb,\KT]=0; & [\Qb,\DT]=[\QBb,\DT]=0; \\

[\QD,\HT]=-2i\gamma\beta^{-1}(\QH-\Qb); &[\QBD,\HT]=-2i\gamma\beta^{-1}(\QBH-\QBb); & \\

[\QD,\KT_0]=2i\gamma\alpha^{-1}(\QK-\Qb); &[\QBD,\KT_0]=2i\gamma\alpha^{-1}(\QBK-\QBb); & \\

[\QD,\DT_0]=0; &[\QBD,\DT_0]=0; &[\QD,\QBD]=4i\gamma\DT_0; \\

[\QK,\HT]=-i\alpha\gamma^{-1}(\QD-\Qb); &[\QBK,\HT]=-i\alpha\gamma^{-1}(\QBD-\QBb); & \\

[\QK,\DT_0]=-i(\QK-\Qb); &[\QBK,\DT_0]=-i(\QBK-\QBb); & \\

[\QK,\KT_0]=0; &[\QBK,\KT_0]=0; &[\QK,\QBK]=2i\alpha\KT_0; \\

[\QH,\QBD]=i\beta\HT+2i\gamma\DT_0-2\beta\gamma H; & 
[\QBH,\QD]=i\beta\HT+2i\gamma\DT_0+2\beta\gamma H; & \\

[\QK,\QBD]=i\alpha\KT_0+2i\gamma\DT_0+2\alpha\gamma K; &
[\QBK,\QD]=i\alpha\KT_0+2i\gamma\DT_0-2\alpha\gamma K; & \\

[\QH,\QBK]=i\beta\HT+i\alpha\KT_0-2\alpha\beta D; & [\QBH,\QK]=i\beta\HT+i\alpha\KT_0+2\alpha\beta D; 
& \\

[\QH,\QBb]=[\QBH,\Qb]=i\beta\HT; & [\QK,\QBb]=[\QBK,\Qb]=i\alpha\KT_0; & \\

[\QD,\QBb]=[\QBD,\Qb]=2i\gamma\DT_0; & & \\

[Q_{\scriptscriptstyle(\ldots)},H]=\beta^{-1}(\Qb-\QH); & [\overline{Q}_{\scriptscriptstyle(\ldots)},H]= 
\beta^{-1}(\QBb-\QBH); & \\

[Q_{\scriptscriptstyle(\ldots)},D_0]=(2\gamma)^{-1}(\Qb-\QD); & 
[\overline{Q}_{\scriptscriptstyle(\ldots)},D_0]=(2\gamma)^{-1}(\QBb-\QBD); & \\

[Q_{\scriptscriptstyle(\ldots)},K_0]=\alpha^{-1}(\Qb-\QK); & 
[\overline{Q}_{\scriptscriptstyle(\ldots)},K_0]=\alpha^{-1}(\QBb-\QBK); & \\

[\HT,H]=0; & [\KT_0,K_0]=0; & [\DT_0,D_0]=0; \\

[\HT,K_0]=[H,\KT_0]=2iD; &
[\HT,D_0]=[H,\DT_0]=iH; & [\DT_0,K_0]=[D_0,\KT_0]=iK. \\

& & \\
\hline
\end{array}
\]

\noindent All other commutators\footnote{The $Q_{(\ldots)}$ appearing in the table can be
any of the following operators: $\Qb$,$\QH$,$\QD$,$\QK$ and the same holds for $\overline{Q}_{(\ldots)}$. 
Obviously 
all commutators are between quantities calculated at the same time.} are zero.

We notice that for our supersymmetric extension we need 14 charges
(see {\bf TABLE~2}) 
in order for the algebra to close, while in the extension of
ref.\scite{FUB} one needs only 8 charges (see {\bf TABLE 1}).

%%%%%%%%%%%%%%%%%%%%%%%%%%%%%%%%%%%%%%%%%%%%%%%%%%%%%%%
%%%%%%%%%%%%%%%%%%

\section{Superconformal Algebras Associated to the Two Extensions.}

A Lie superalgebra\scite{KAC} is an algebra made of even $E_{n}$ and
odd $O_{\alpha}$ generators whose graded commutators look like:

\be
\label{eq:venticinque}
[E_{m},E_{n}]=F^{p}_{mn}E_{p};~~
[E_{m},O_{\alpha}]=G^{\beta}_{m\alpha}O_{\beta};~~
[O_{\alpha},O_{\beta}]=C^{m}_{\alpha\beta}E_{m};
\ee 

\noindent and where the structure constants $F^{p}_{mn},G^{\beta}_{m,\alpha},
C^{m}_{\alpha,\beta}$ satisfy generalized Jacobi identities.

The second relation of eq.(25) is usually interpreted by saying that
the even part of the algebra has a representation on the odd part. 
This is clear if we consider the odd part as a vector space and that
the even part acts on this vector space via the graded
commutators. 

For superconformal algebras the usual folklore says that the even part of the
algebra has his conformal subalgebra represented spinorially
on the odd part. This is true only in a relativistic setting
and it is not always the case in a non-relativistic one. We will now
analyze both the case of ref.\scite{FUB} and ours.

Let us start from the superalgebra of  ref.\scite{FUB}  which is given in 
eqs.(14)--(16).
The even part of this superalgebra ${\mathcal G}_{0}$ can be 
organized in an $SO(2,1)$ form as follows:

\[
{\mathcal G}_{0}: \left\{ 
\begin{array}{l}
B_{1}=\displaystyle{1\over 2}\left[ {K\over a}-aH \right] \\
B_{2}= D \\
J_{3}=\displaystyle{1\over 2}\left[ {K\over a}+a H \right]
\end{array}
\right.
\]

\noindent where $a$ is the same parameter introduced in \cite{DFF} with dimension of
time.

On the other side the odd part ${\mathcal G}_{1}$ is:

\[
{\mathcal G}_{1}: \left\{ \begin{array}{l}
Q\\
Q^{\dag}\\
S\\
S^{\dag}
\end{array}
\right.
\]

\noindent It is easy to work out, using the results of {\bf TABLE 1}, the action
of the ${\mathcal G}_{0}$  on ${\mathcal G}_{1}$. The result
is summarized in  table {\bf 6} of ref.\scite{PINO}.

Considering the odd part as a vector
space, let us build the following 4 ``vectors": 
\bea
\label{eq:ventisei}
&&|q\rangle\equiv Q+Q^{\dag};~~~~~
|p\rangle  \equiv  S-S^{\dag};\\
\label{eq:ventisette}
&&|r\rangle\equiv Q-Q^{\dag};~~~~~
|s\rangle\equiv  S+S^{\dag}.
\eea

\noindent Next let us take the Casimir operator
of the algebra ${\mathcal G}_{0}$ which is~
${\mathcal C}=B_{1}^{2}+B_{2}^{2}-J_{3}^{2}$
and apply it to the state $|q\rangle$:

\be
\label{eq:ventotto}
{\mathcal C}|q\rangle=[B_{1},[B_{1},Q+Q^{\dag}]]+[B_{2},[B_{2},Q+Q^{\dag}]]-
[J_{3},[J_{3},Q+Q^{\dag}]]
=-{3\over 4}|q\rangle.
\ee

\noindent The same happens for the state $|p\rangle$.
The factor ${3\over 4}=-{1\over 2}({1\over 2}+1)$ above indicates that the
$(|q\rangle,|p\rangle)$ space carries a spinorial representation. 
It is possible to prove the same for the other two vectors.

Let us now turn the same crank for our supersymmetric extension
of conformal mechanics. Looking at the {\bf TABLE 2} of our operators,
we can organize the even part ${\mathcal G}_{0}$, as follows:
\vskip .5 cm
\begin{center}
{\bf TABLE 4 (${\mathcal G}_{0}$)}
\end{center}
\[
\begin{array}{|lcl|}
\hline 

&&\\
B_{1}=\displaystyle{1\over 2}\left( {\KT\over a}-a\HT\right); && P_{1}=2D;\\

B_{2}=\displaystyle\DT; && P_{2}=\displaystyle aH-{K\over a};\\

J_{3}=\displaystyle{1\over 2}\left( {\KT\over a}+a\HT\right); && P_{0}=\displaystyle aH+{K\over a}.\\
&&\\
\hline
\end{array}
\]
\vskip .5 cm
\noindent The LHS is the usual $SO(2,1)$ while the RHS is formed by three translations
because they commute among themselves. So the overall algebra is the Euclidean
group $E(2,1)$.

The odd part of our superalgebra is made of 8 operators (see {\bf TABLE 2}).
\noindent As we did before  for the model of\scite{FUB},
we will now evaluate for our model the action of ${\mathcal G}_{0}$ 
on the odd part. The result is summarized in table {\bf 9} of ref.\scite{PINO}.
It is easy\scite{PINO} to realize from  that table that  the following three vectors
(where for simplicity we have made the choice 
$\displaystyle a=\sqrt{\frac{\beta}{\alpha}}$ 
and $\displaystyle\eta\equiv\frac{\gamma}{\sqrt{\alpha\beta}}$):

\be
\left\{
\begin{array}{l}
|q_{\scriptscriptstyle H}\rangle = (\QH-\Qb)-(\QBH-\QBb)\\
|q_{\scriptscriptstyle K}\rangle = (\QK-\Qb)-(\QBK-\QBb)\\
|q_{\scriptscriptstyle D}\rangle = \eta^{-1}[(\QD-\Qb)-(\QBD-\QBb)]
\end{array}
\right.
\ee
\noindent
make an irreducible representation of the conformal subalgebra. In fact 
one easily\scite{PINO} obtains:
\bea
\label{eq:trenta}
&&B_{1}|q_{\scriptscriptstyle H}\rangle=-{i\over 2}
|q_{\scriptscriptstyle D}\rangle;~~~~
B_{2}|q_{\scriptscriptstyle H}\rangle=-i|q_{\scriptscriptstyle H}\rangle;~~~~
J_{3}|q_{\scriptscriptstyle H}\rangle=-{i\over 2}
|q_{\scriptscriptstyle D}\rangle\nonumber\\
&&B_{1}|q_{\scriptscriptstyle K}\rangle=-{i\over 2}
|q_{\scriptscriptstyle D}\rangle;~~~~
B_{2}|q_{\scriptscriptstyle K}\rangle=i|q_{\scriptscriptstyle K}\rangle;~~~~~~
J_{3}|q_{\scriptscriptstyle K}\rangle={i\over 2}|q_{\scriptscriptstyle
D}\rangle\\
&&B_{1}|q_{\scriptscriptstyle D}\rangle= 
-i(|q_{\scriptscriptstyle H}\rangle
+|q_{\scriptscriptstyle K}\rangle);~~~~
B_{2}|q_{\scriptscriptstyle D}\rangle=0;~~~~
J_{3}|q_{\scriptscriptstyle D}\rangle=i(|q_{\scriptscriptstyle H}\rangle-
|q_{\scriptscriptstyle K}\rangle)\nonumber.
\eea

\noindent The Casimir operator is given, as before, by: 
${\mathcal C}=B_{1}^{2}+B_{2}^{2}-J_{3}^{2}$ but we must remember to use, for 
$B_1$, $B_2$ and $J_3$, the operators contained
in {\bf TABLE~4}. It is then easy to check that:
${\mathcal C}|q_{\scriptscriptstyle H}\rangle=-2
|q_{\scriptscriptstyle H}\rangle$. The same we get for the other two vectors $|q_{\scriptscriptstyle 
K}\rangle,
|q_{\scriptscriptstyle D}\rangle$, so the eigenvalue in the equation above is\break ${\mathcal C}=-2=-1(1+1)$ 
and this indicates that those vectors make a spin-1 representation.
It is also easy to prove that there is another spin-1 representation
and two spin-0. The details can be found in ref.\scite{PINO}.

We wanted to present this analysis in order to underline a further difference
between our supersymmetric extension and the one of\scite{FUB}
whose odd part ${\mathcal G}_{1}$, as we showed before, carries
two spin one-half representations.
%%%%%%%%%%%%%%%%%%%%%%%%%%%%%%%%%%%%%%%%%%%%%%%%%%%%%%%
%%%%%%%%%%%%%%%%%%%%%%%

\section{Exact Solution of the Model and Its Superspace Formulation.}

\hspace{.4 cm}We will now present a superspace formulation of our
model like the authors of ref.\cite{FUB} did for theirs.
We have to enlarge the  ``base space" $(t)$ to a superspace $(t,\theta,{\bar\theta})$
where $(\theta, {\bar\theta})$ are Grassmannian partners of $(t)$.
It is then easy to put all the variables \quattrova~in a single superfield
$\Phi$ defined as follows:
\be
\label{eq:trentuno}
\Phi^{a}(t,\theta,{\bar \theta})=\phi^{a}(t)+\theta c^{a}(t)+
{\bar \theta}~\omega^{ab}{\bar c}_{b}(t)+i{\bar\theta}\theta~\omega^{ab}\lambda_{b}(t).
\ee

\noindent This superfield had already been introduced in ref.\cite{ENNIO}. It is
a scalar field under the supersymmetry transformations of the system.
It is a simple exercise to find the expansion of any function $F(\Phi^{a})$ 
of the superfields in terms of $\theta,{\bar\theta}$. For example, 
choosing as function the Hamiltonian $H$ of a system, we get:
\be
\label{eq:trentadue}
H(\Phi^{a})=H(\phi)+\theta N_{\scriptscriptstyle H}-{\bar\theta}~
{\overline N}_{\scriptscriptstyle H}+i\theta{\bar\theta}~\HT.
\ee
\noindent From eq.(\ref{eq:trentadue}) it is easy to prove that:
\be
\label{eq:trentatre}
i\int H(\Phi)~d\theta d{\bar\theta}=\HT.
\ee

\noindent Here we immediately notice a crucial difference with the supersymmetric
QM model of ref.\cite{FUB}. In the language of superfields (see the second of ref.\scite{FUB})
those authors obtained the supersymmetric potential
of their Hamiltonian  by inserting
the superfield into the superpotential (which is given by
eq.({\ref{eq:tredici})) and integrating in  something
like $\theta,{\bar\theta}$, while we get the potential part
of our supersymmetric Hamiltonian by inserting the superfield into the
normal potential of the conformal mechanical model given in (\ref{eq:nove}).

The space \quattrova  somehow can be considered 
as a {\it target} space whose {\it base} space is the superspace 
$(t,\theta,{\bar\theta})$. The action of the various charges listed in our
{\bf TABLE~2} is on the target-space variables but we can consider it as
induced by some transformations on the base-space. If we collectively
indicate  the charges acting on \quattrova  as $\Omega$, we shall indicate
the generators of the corresponding transformations on the base space
as ${\widehat \Omega}$. The relation between the two is the following:
\be
\label{eq:trentaquattro}
\delta\Phi^{a}=-\varepsilon {\widehat \Omega}\Phi^{a}~~~\mbox{where}~~~
\delta\Phi^{a}=[\varepsilon\Omega,\Phi^{a}],
\ee
with $\varepsilon$ the commuting or anticommuting infinitesimal parameter
of our transformations and $[(.),(.)]$
the graded commutators of our formalism.

Using the relations above it is easy to work out the superspace representation
of the operators of eq.(18). They are:
\bea
\label{eq:trentacinque}
&&\QCb=-\partial_{\theta};~~~~~
\QCBb = \partial_{\bar\theta};~~~~~
{\widehat Q}_{g} ={\bar\theta}\partial_{\bar\theta}-
\theta\partial_{\theta}  \\
&&{\widehat C}={\bar\theta}\partial_{\theta};~~~~
{\widehat {\overline C}} = \theta\partial_{\bar\theta};~~~~
\widehat{N}_H = {\bar\theta}\partial_{t};~~~~
\widehat{\overline{N}}_H =\theta\partial_{t}\nonumber
\eea
\noindent Via the charges above it is easy to write down also the
supersymmetric charges of eq.(\ref{eq:diciannove}):\break
$\QCH=-{\partial}_{\theta}-\beta~{\bar\theta}\partial_{t};~~
\QCBH={\partial}_{\bar\theta}+\beta~{\theta}\partial_{t}$.

\noindent Proceeding in the same way, via the relation (\ref{eq:trentaquattro}),
it is a long but easy procedure to give a superspace
representation to the charges $\QD,\QBD,\QK,\QBK$ of
eqs. (\ref{eq:ventotto})--(\ref{eq:trentuno}).The result\scite{PINO} is:

\bea
\label{eq:trentasei}
&&\HCT=\displaystyle i\frac{\partial}{\partial t};~~~~~~
\DCT=\displaystyle i t\frac{\partial}{\partial t}-\frac{i}{2}\sigma_3;~~~~~
\KCT=\displaystyle it^2\frac{\partial}{\partial
t}-it\sigma_3-i\sigma_-;\nonumber\\
&&\widehat{Q}^t_{\scriptscriptstyle
D}=\displaystyle-\frac{\partial}{\partial\theta}-
2\gamma~{\bar\theta}t\frac{\partial}{\partial
t}+\gamma~{\bar\theta}\sigma_3;~~~~~
\widehat{\overline{Q}}^t_{\scriptscriptstyle
D}=\displaystyle\frac{\partial}{\partial\bar{\theta}}+2\gamma~\theta
t\frac{\partial}{\partial t}-\gamma{\theta}\sigma_3;\\
&&\widehat{Q}^t_{\scriptscriptstyle
K}=\displaystyle-\frac{\partial}{\partial\theta}-\alpha~{\bar\theta}t^2
\frac{\partial}{\partial t}+\alpha~t{\bar\theta}\sigma_3+\alpha~
{\bar\theta}\sigma_{-};~~~~
\widehat{\overline{Q}}^t_{\scriptscriptstyle
K}=\displaystyle\frac{\partial}{\partial\bar{\theta}}+\alpha\theta~
t^2\frac{\partial}{\partial t}-\alpha t~\theta\sigma_3-\alpha\theta\sigma_{-};
\nonumber\\
&&\hat{H}=\displaystyle\bar{\theta}\theta\frac{\partial}{\partial t};~~~~
\hat{D}=\displaystyle\bar{\theta}\theta~(t\frac{\partial}{\partial
t}-\frac{1}{2}\sigma_3);~~~~
\hat{K}=\displaystyle\bar{\theta}\theta~(t^2\frac{\partial}{\partial
t}-t\sigma_3-\sigma_{-}).\nonumber
\eea

\noindent In the previous expressions the $\sigma_{3}$ and $\sigma_{-}$ are the Pauli
matrices while the  index $(\cdot)^{t}$
indicates an explicit dependence
on $t$ which appears for the following reasons. Let us go back to
relation (6): 
\bea
&&H = H_{0};\nonumber\\
&&D = t H+D_{0};\nonumber\\
&&K = t^{2}H+2tD_{0}+K_{0};\nonumber
\eea
\noindent from which we get:
\bea
&&\DT = t\HT+\DT_{0};\nonumber\\
&&\KT =  t^{2}\HT+2t\DT_{0}+\KT_{0}.
\eea
\noindent The ``square roots" of these operators
will depend explicitly on the time $t$ and they are those listed in eqs.(36).
For more details about their derivation we invite the reader to consult
reference \scite{PINO}.

Let us now turn to the solution of our model.
The original conformal mechanical model was solved exactly
in eq.(2.35) of reference \cite{DFF}. The solution was given by the relation:
\be
\label{eq:trentotto}
q^{2}(t)=2t^{2}H-4t D_{0}+2K_{0}.
\ee

\noindent As $(H,D_{0},K_{0})$ are constants of motion, once their values are assigned we
stick them in eq. (\ref{eq:trentotto}), and we get a relation between ``$q$" 
(on the LHS of 
(38) ) and ``$t$" on the RHS. This is the solution of the equation
of motion with ``initial conditions" given by the values we assign
to the constants of motion $(H,D_{0},K_{0})$. The reader may object
that we should give only 
two constant values (corresponding to the initial conditions $(q(0),{\dot q}(0))$) and not three.  
Actually the three values assigned to $(H,D_{0},K_{0})$ are not arbitrary because, as it was
proven in eq.(2-36) of ref.\scite{DFF}, these three quantities are linked
by a constraint: $\left(HK_{0}-D_{0}^{2}\right)={g\over 4}$
where ``$g$" is the coupling which entered the original Hamiltonian (see eq.(1)
of the present paper). Having one constraint among the three constants of motion brings them
down to two. 

What we want to do here is to see if a relation
analogous to (38) exists also for our supersymmetric extension
or in general if the supersymmetric system can be solved exactly.
The answer is {\it yes} and it is based on a very simple trick.
Let us first remember how $\HT$ and $H$ are related:~~$
i\int H(\Phi)~d\theta d{\bar\theta}=\HT$.
The same relation holds for $\DT_{0}$ and $\KT_{0}$ with respect to $D_{0}$
and $K_{0}$:
\be
\label{eq:trentanove}
i\int D_{0}(\Phi)~d\theta d{\bar\theta}=  \DT_{0};~~~~
i\int K_{0}(\Phi)~d\theta d{\bar\theta}=  \KT_{0}.
\ee

\noindent Of course the same kind of relations holds also for the explicitly time-dependent
quantities:\break
$i\int D(\Phi)~d\theta d{\bar\theta}= \DT;~~
i\int K(\Phi)~d\theta d{\bar\theta}=  \KT$.

\noindent Let us now build the following quantity:
\be
\label{eq:quaranta}
2t^2H(\Phi)-4tD(\Phi)+2K(\Phi);
\ee

\noindent this is functionally the RHS of eq.(38) with
the superfield $\Phi^{a}$ replacing the normal phase-space variable
$\phi^{a}$. It is then clear that the following relation holds:
\be
\label{eq:sessantuno}
\left(\Phi^{q}\right)^{2}=2t^2H(\Phi)-4tD(\Phi)+2K(\Phi).
\ee

\noindent The reason it holds is because, in the proof\scite{DFF} 
of the analogous one in $q$-space, the only thing the authors used
\scite{DFF}  was the functional form of the $(H,D,K)$.
So that relation holds irrespective
of the arguments, $\phi$ or $\Phi$, which enter our functions provided that
the functional form of them remains the same. From the form of the 
superfields it is then easy to expand in $\theta$
$\bar\theta$ the RHS and LHS of the relation (40) and get four relations, 
one for each of the variables $(q,c^{q},{\bar c}_{p},\lambda_{p})$,
entering our formalism. These relations solves the model completly.
For more details we refer the reader to ref.\scite{PINO}.

\end{document}